# Nonextensive Thermostatistical Investigation of

# The Blackbody Radiation

**Fevzi Büyükkılıç[a†], İsmail Sökmen[b], Doğan Demirhan[a]**

[a] *Department of Physics, Faculty of Sciences, Ege University, 35100 Bornova İzmir, Turkey*

[b] *Department of Physics, Faculty of Arts&Sciences, Dokuz Eylül University, Alsancak İzmir, Turkey*

**Abstract**

Thermodynamical quantities of the blackbody radiation, such as free energy, entropy, total radiation energy, specific heat are calculated within the Tsallis thermostatistics where factorization method is incorparated. It is shown that basic thermodynamical relation of the blackbody radiation is form invariant with respect to nonextensivity entropic index q. Furthermore, the nonextensive thermodynamical quantities related to the blackbody radiation is seperately be obtained in terms of q and the standard thermodynamical quantities of the blackbody radiation. It is indicated that the formulation may give a way to determine the q which determines the degree of the nonextensivity that is the one of the aims of the present study.

## 1.Introduction

It has been understood that extensive (additive) Boltzmann-Gibbs(BG) thermostatistics fails to study nonextensive physical systems where long range interactions or long-range microscopic memory is involved or, the system evolves in a (multi)fractal space-time. Thus, the standard statistical mechanic is not universal and is valid for extensive systems. A generalized formalism is proposed by C.Tsallis which describe the features that nonextensive physical systems exhibit.[1,2,3].

This generalization relies on a new entropic form of the entropy that is inspired from (multi)fractals

$$Sq = k_B \frac{1 - \sum_{i=1}^{W} p_i^q}{q-1} \qquad (q \in R)$$

[†] Correspondence address: Department of Physics, Faculty of Science, Ege University, 35100 Bornova/İzmir/Turkey    e-mail:fevzi@fenfak.ege.edu.tr

where $k_B$ is a positive constant and W is the total number of microscopic accessible states of the system. (for the q<0 case, those probabilities which are not positive should be excluded). This expression recovers the well-known standard Shannon entropy in the limit q→1

$$S_1 = -k_B \sum_{i=1}^{W} p_i \ln p_i.$$

The nonextensivity entropic index q, related to and determined by the microscopic dynamics of the system , charecterizes the degree of nonextensivity.

Recently, the generalized statistical mechanics has succesfully been applied to investigate physical systems which exhibit nonextensive features. Amongst them stellar polytrops [3], Levy-like anomalous diffusions [4], two dimensional turbulance[5], solar neutrino problem[6] velocity distribution of galaxy clusters [7], Cosmic background radiation [8,9] and correlated themes [10], linear response theory [11], thermalization of electron-phonon system [12], and low dimensional dissipative systems [13] could be enumareted

It should be remarked that what makes $S_q$ favourable is that it has, with regard to {$p_i$}, definite concavity property for all values of q [ 14].
In order to have sensible results which are exact or approximate ,some techniques of calculations have been formulated in Tsallis generalized statistical mechanics formalism (TT). These are (1-q) expansion [8], factorization method for quantal distribution functions [15], perturbative expansion [16], variational methods [16,17], semiclassical approximation [18], Feynman's path integrals method in nonextensive physics [19], and Green functions [20].

In a recent paper, two of the authors FB and DD, regarding the particles of fermion and bosons as g-ons, has obtained a distribution function which unifies the nonextensive distribution function of quantum gases; i.e. bose and fermi gases [21].

In this paper, however, we have revisited the blackbody radiation where it is investigated by using factorization method within TT [22]. Here, our motivation is to redrive the thermodynamical quantities, by incorporating factorization approximation in the context of nonextensive thermostatistics, moreover, to reassess the obtained results with the standard ones. We believe that our approach will help to understand the thermodynamical difference between extensive and a nonextensive blackbody radiation  where nonextensive thermodynamical quantities is seperately revealed in terms of entropic index q as well as extensive standard ones.

## 2. The Generalized Law of Planck

We would like to redrive  the generalized Law of Planck by introducing the newly drived distribution function for quantum gases which unifies Bose-Einstein and Fermi-Dirac distribution. Electromagnetic radiation which is in thermal equilibrium and which may be regarded as a gas of photons is named as blackbody radiation. Photons do not interact with one another, so that the photon gas is an ideal quantum gas. The mechanism by which equilibrium can be established consists in the absorption and emission of the photon gas: the number of photons N in it is a variable, and not a given constant is in ordinary gas, from the requirement that the free energy of the gas should be a minimum, the chemical potential of the photon gas µ is zero. Two of the authors unified the distribution function of the bose and the fermi gas within the fractal and fractional approach. According to the formula which is derived the most probable distribution of the $N_q$ g-ons over the single state is

$$\bar{n}(g,q) = \frac{1}{\left[1-(1-q)\beta(\epsilon_k - \mu)\right]^{\frac{1}{q-1}} + 2g - 1} \tag{1}$$

where g is a statistics number, $\beta = \frac{1}{k_B T}$, $k_B$ is Boltzmann constant, µ the chemical potential. It can be verified that in the $q \to 1$ limit for g=0 Bose-Einstein and for g=1 Fermi-Dirac distributions are recovered.

Eq.(1) may also lead to a generalized Planck distribution for µ=0 and g=0 since the angular momentum of photons is unity therefore they behave as bosons. Standard Planck Distribution Law, however, is recovered for µ=0, g=0 and q=1

The distribution of photons among the various quantum states with energies $\epsilon_k = \hbar\omega_k$ is given by the generalized Planck distribution formula (1) with µ=0; and g=0, that is,

$$\bar{n}_k^q = \frac{1}{\left[1-(1-q)\beta\hbar\omega\right]^{\frac{1}{q-1}} - 1} . \tag{2}$$

For a sufficiently large volume, one can pass from the discrete to the continous distribution of eigen frequencies of radiation. Thus, the density of states of the photons, with frequncies between ω and ω+dω in a given volume V [23] is

$$D(\omega) = \frac{V\omega^2 d\omega}{\pi^2 c^3} . \tag{3}$$

The number of photons in the frequency interval $d\omega$ is obtained by multiplying $\bar{n}_k^q$ with the generalized distribution of Planck which is given by Eq.(3):

$$dN_q(\omega) = \frac{V}{\pi^2 c^3} \frac{\omega^2 d\omega}{\left[1-(1-q)\beta\hbar\omega\right]^{\frac{1}{q-1}} - 1} \tag{4}$$

or
$$\frac{dn(\omega)}{d\omega} = \frac{1}{\pi^2 c^3} \frac{\omega^2}{\left[1-(1-q)\beta\hbar\omega\right]^{\frac{1}{q-1}} - 1} \quad (5)$$

where $n = N/V$ is the spatial density of photons

Multiplying Eq.(4) by $\hbar\omega$ leads to the radiation energy for this frequency interval:

$$dU_q(\omega) = \frac{V\hbar}{\pi^2 c^3} \frac{\omega^3 d\omega}{\left[1-(1-q)\beta\hbar\omega\right]^{\frac{1}{q-1}} - 1}. \quad (6)$$

Quite analogously, one finds the spatial energy density $u = \frac{U}{V}$ per frequency interval:

$$\frac{du_q(\omega)}{d\omega} = \hbar\omega \frac{dn_q}{d\omega} = \frac{\hbar}{\pi^2 c^3} \frac{\omega^3}{\left[1-(1-q)\beta\hbar\omega\right]^{\frac{1}{q-1}} - 1}. \quad (7)$$

The rate R at which photons with frequency ω leave the cavity through the small outlet is

$$R_q(\omega) = \frac{c}{4} \frac{dn_q(\omega)}{d\omega} = \frac{1}{4\pi^2 c^3} \frac{\omega^3}{\left[1-(1-q)\beta\hbar\omega\right]^{\frac{1}{q-1}} - 1} \quad (8)$$

The energy flux per unit area of the hole per unit time and frequency interval is thus

$$\frac{d^3 E_q}{dFd\omega dt} = \hbar\omega R_q(\omega). \quad (9)$$

Since the energy per unit time just yields the radiative power $P_q$, Eq.(9) is, of course, identical with radiative power per hole area and frequency, i.e., with spectral density

$$Q_q^s(\omega, T) = \hbar\omega R_q(\omega) = \frac{\hbar}{4\pi^2 c^2} \frac{\omega^2}{\left[1-(1-q)\beta\hbar\omega\right]^{\frac{1}{q-1}} - 1} \quad (9\,a)$$

This is the generalized radiation law of Planck for the spectral radiation density of a a blackbody in thermal equilibrium within the dilute gas approximation. Or in terms of the wavelength $\lambda = 2\pi c/\omega$, it becomes[21] the genarilized Law of Planc:

$$dU_\lambda = \frac{16\pi^2 c\hbar V}{\lambda^5} \frac{d\lambda}{\left[1-(1-q)\frac{2\pi\hbar}{k_B T}\right]^{\frac{1}{1-q}} - 1}. \quad (9\,b)$$

## 3. Generalized Thermodynamics of a Black-body Radiation

Let us calculate generalized thermodynamic quantities,namely,the free enrgy,the entropy, the total radiation energy,the epecific heat, the pressure of a blackbody radiation within the factorization approximation.

The free energy for the chemical potential µ=0 ;  $F_q = -k_B T \ln Z_q$

$$F_q = -kT \sum_{i=1}^{\infty} \ln \left( \sum_{n=0}^{\infty} [1-(1-q)n\beta\hbar\omega_i]^{\frac{1}{1-q}} \right) \qquad (10)$$

or

$$F_q = -k_B T \sum_{i=1}^{\infty} \ln \left( \sum_{n=0}^{\infty} \ln \frac{1}{[1-(1-q)n\beta\hbar\omega_i]^{\frac{1}{1-q}}} \right). \qquad (11)$$

By means of desity of the quantum states for the interval dω which is given by Eq.(3), one passes from the summation to integration in Eq.(11) thus we obtain

$$F_q = -\frac{V}{\pi^2 c^3} \int \omega^2 \ln \left( 1-[1-(1-q)n\beta\hbar\omega_i]^{\frac{1}{1-q}} \right) d\omega. \qquad (12)$$

With the new variable of integration $x = \frac{\hbar\omega}{k_B T}$, integration by parts gives , in the (1-q)x→0 limit;i.e.within the factorization approximation which is developed in ref[15]

$$F_q == -V \frac{(kT)^4}{3\pi^2 \hbar^3 c^3} I_q \qquad (13)$$

where

$$I_q = \int_0^{\infty} \frac{x^3}{[1-(1-q)x]^{\frac{1}{1-q}} -1} dx \qquad for \quad \begin{matrix} q>1 \\ q<1 \end{matrix} \qquad (14)$$

The value of integral for both q>1 or q<1 cases is obtained as

$$I_q = \frac{\pi^4}{15} \frac{1}{(4-3q)(3-2q)(2-q)} \qquad (15)$$

which appears in Eq.(A24)of the AppendixA. Due to singularities nonextensivity index q can not have the values 4/3,3/2, and 2.The largest value for nonextensivity index q can be 4/3.It is obtained 5/4 in ref [] where the integration is limited to a cut-off frequency.

Substitution of Eq.(15) into Eq.(13) gives for the free energy

$$F_q = -\frac{4V}{3c}\sigma_q T^4 \tag{16}$$

where

$$\sigma_q = \frac{\sigma}{(4-3q)(3-2q)(2-q)} \tag{16 a}$$

is the q-dependent and

$$\sigma = \frac{\pi^2 k^4}{60\hbar^3 c^2}$$

is standard Stefan-Boltzmann constant.So $\sigma_q$ is independent of the temperature.

Serial representation of Eq.(16a) can be realized by Taylor expansion with respect to entropic index about q=1. Thus,

$$\sigma_q = \sigma(1 + 6(q-1) + 25(q-1)^2 + O(q-1)^3). \tag{16b}$$

In the $q \to 1$ limit, $\sigma_q$ become the conventional Stefan-Boltzmann constant.

On the other hand ,the entropy of the blackbody radiation,naturally , is found to be q-dependent

$$S_q = -\left(\frac{\partial F_q}{\partial T}\right) = \frac{16V}{3c}\sigma_q T^3. \tag{17}$$

where $\sigma_q$ is given in Eq.(16a).

The total radiation energy is found to be

$$U_q = F_q + TS_q. \tag{18}$$

It is seen that Eq.(18) is form invariant.

Substitution of Eqs.(16) and (17) into Eq.(18) leads to

$$U_q = \frac{4V}{c}\sigma_q T^4 \tag{19}$$

or

$$U_q = -3F_q. \tag{20}$$

On the other hand, the direct intgration of the Eq.(6) without a upper limit of the integral and a serial representation of the distribution function [] with respect to ω leads to the Eq.(19)

$$\int_0^\infty dU_q(\omega)\, d\omega = \frac{V\hbar}{\pi^2 c^3} \int_0^\infty \frac{\omega^3 d\omega}{[1-(1-q)\beta\hbar\omega]^{\frac{1}{q-1}} - 1}. \tag{21}$$

With the new variable of integration $x = \frac{\hbar\omega}{kT}$, Eq.(21) reads

$$U_q = \frac{V\hbar}{\pi^2 (c\hbar)^3 \beta^4} \int_0^\infty \frac{x^3 dx}{[1-(1-q)x]^{\frac{1}{q-1}} - 1} \tag{22}$$

or

$$U_q = \frac{V(kT)^4}{\pi^2 (c\hbar)^3} I_q. \tag{23}$$

The value of the integral is given in Eq.(A10) of the Appendix. Therefore

$$I_q(3) = \frac{I}{(4-3q)(3-2q)(2-q)} \tag{24}$$

where I is given as:

$$I = \frac{\pi^2}{15} \tag{25}$$

Substitution of Eq.(A 10) into Eq.(23) leads to

$$U_q = \frac{V(kT)^4}{(c\hbar)^3} \frac{\pi^2}{15} \frac{1}{(4-3q)(3-2q)(2-q)} \tag{26}$$

or, in a more compact form;

$$U_q = \frac{4V}{c} \sigma_q \tag{27}$$

where $\sigma_q$ is the same with the one which is given explicitly by Eq.(16 a).

The specific heat of radiation is

$$C_V^q = \frac{16V}{c} \sigma_q T^3. \tag{28}$$

The pressure is

$$p_q = -\left(\frac{\partial F}{\partial V}\right)_T = \frac{4}{3c} \sigma_q T^4 \tag{29}$$

or

$$p_q V = \frac{1}{3} E_q. \tag{30}$$

Eq.(30) for a photon gas is a direct consequence of the linear ralation $\epsilon = cp$ between the energy and momentum of a photon.

## 4. The Total Number of photons of a Black-body Radiation

By taking the integration of the Eq.(4) the total number of photons in a generalized blackbody radiation is

$$\int_0^\infty dN_q(\omega)d\omega = \frac{V}{\pi^2 c^3} \int_0^\infty \frac{\omega^2 d\omega}{[1-(1-q)\beta\hbar\omega]^{\frac{1}{q-1}} - 1} \tag{31}$$

with the new variable of integration $x = \frac{\hbar\omega}{kT}$,

$$N_q = \frac{VT^3}{\pi^2 c^3 \hbar^3} I_q \tag{32}$$

where

$$I_q(2) = \int_0^\infty \frac{x^2}{[1-(1-q)x]^{\frac{1}{q-1}} - 1} dx. \tag{33}$$

The value of the integral is given in Eq.(A) of the Appendix. Thus,

$$I_q(2) = \frac{I}{(3-2q)(2-q)} \quad for \quad \begin{array}{c} q<1 \\ q>1 \end{array} \tag{34}$$

where

$$I = \int_0^\infty \frac{x^2 dx}{e^x - 1} = \Gamma(3)\xi(3). \tag{35}$$

Substitution of Eq.(33) into Eq.(32) while taking into account the Eqs.(34) and Eq.(35) ;one finds

$$N_q = \frac{N}{(3-2q)(2-q)} \tag{36}$$

where $$N = \frac{V}{\pi^2}\left(\frac{T}{\hbar c}\right)^3 \Gamma(3)\xi(3) \tag{37}$$

is the standard number of the photons.

Serial representation of Eq.(36) can be realized by Taylor expansion with respect to entropic index about q=1. Thus

$$N_q = N(1 + 3(q-1) + 7(q-1)^2 + O^3(q-1))). \tag{38}$$

İn the $q \to 1$ limit $N_q$ become conventional N in Eq.(37).

## 5. Conclusions

In this study the blackbody radiation problem is revisited and Generalized The Planck Law is rederived The thermodynamical quantities of the blackbody; i.e. the Free energy, the entropy, the total radiation energy, the specific heat and the pressure of a blackbody radiation are calculated within the Tsallis thermostatistics in which factorization method is involved.

It is seen that, basic thermodynamical relations related to blackbody radiation are form invariant with respect to nonextensivity entropic index q which determines the the degree of nonextensivity . It should be pointed out that the nonextensive quantities related to the blackbody which are presented have been seperately written in terms of entropic index q and standard corresponding thermodynamical quantities.. We hope that our approach may well lead to determine the entropic index q, so that the ultimate aim of the study can be  achieved.

## Appendix A

In this appendix the calculation of the one of the two integrals which appeared in the text are presented.The integral $I_q(3)$ which has appeared in Eq.(10) of Helmholtz free energy and Eq.(22) of total energy $U_q$ is given by

$$I_q(3) = \int_0^\infty dx \frac{x^3}{[1-(1-q)x]^{\frac{1}{q-1}} - 1} \qquad q<1 \ case. \tag{A1}$$

The definition of the gamma function is

$$\Gamma(\alpha) = \int_0^\infty t^{\alpha-1} e^{-t} dt; \qquad \alpha > 0. \tag{A2}$$

By substituting $t = [1-(1-q)x]\upsilon$ in Eq.(12) one finds

$$[1-(1-q)x]^{\frac{1}{q-1}} = \frac{1}{\Gamma\left(\frac{1}{1-q}\right)} \int_0^\infty \upsilon^{\frac{1}{q-1}-1} e^{-\upsilon} e^{\upsilon(1-q)x} d\upsilon \tag{A3}$$

where $\alpha = \frac{1}{1-q}$ is obtained and the condition for $\alpha > 0$ is fulfilled since q<1 .

On the other hand from the definition of the gamma function it can be written:

$$1 = \frac{1}{\Gamma\left(\frac{1}{1-q}\right)} \int_0^\infty v^{\frac{1}{1-q}-1} e^{-v} dv. \tag{A4}$$

If one substracts Eq.(A3) from Eq.(A4) one achieves

$$[1-(1-q)x]^{\frac{1}{q-1}} - 1 = \frac{1}{\Gamma\left(\frac{1}{1-q}\right)} \int_0^\infty v^{\frac{1}{q-1}-1} e^{-v} \left(e^{v(1-q)x} - 1\right) dv. \tag{A5}$$

Substituing Eq.(A5) into Eq.(A1), after some rearrangements, leads to

$$I_q(3) = \frac{1}{\Gamma\left(\frac{1}{1-q}\right) \int_0^\infty dv\, v^{\frac{1}{q-1}-1} e^{-v}} \int_0^\infty dx \frac{x^3}{e^{v(1-q)x} - 1} \tag{A6}$$

Or in terms of $y = v(1-q)x$ Eq.(A6) takes the form

$$I_q(3) = \frac{\Gamma\left(\frac{1}{1-q}\right)}{(1-q)^4 \int_0^\infty dv\, v^{\frac{1}{1-q}+3} e^{-v}} \int_o^\infty dy \frac{y^3}{e^y - 1} \tag{A7}$$

or

$$I_q = \frac{\pi^4}{15} \frac{\Gamma\left(\frac{1}{1-q}\right)}{(1-q)^4 \Gamma\left(\frac{1}{1-q}+4\right)} \tag{A8}$$

where

$$\int_o^\infty dy \frac{y^3}{e^y - 1} = \frac{(2\pi)^4}{4.2} B_2 = \frac{\pi^4}{15} \tag{A9a}$$

where $B_2 = \frac{1}{30}$ ia a Bernoulli number[24].

$$\int_0^\infty dv\, v^{\frac{1}{1-q}+3} e^{-v} = \Gamma\left(\frac{1}{1-q}+4\right) \tag{A9b}$$

are used.

On the other hand, in view of the property of the gamma function one can write

$$\Gamma(z+1) = z\Gamma(z). \tag{A10a}$$

Thus,
$$\Gamma\left(\frac{1}{1-q}+4\right)=\Gamma\left(\frac{1}{1-q}+3\right)\Gamma\left(\frac{1}{1-q}+2\right)\Gamma\left(\frac{1}{1-q}+1\right)\Gamma\left(\frac{1}{1-q}\right). \tag{A10b}$$

When Eq.(A10b) is substituted in Eq.(A8) one obtains

$$I_q(3)=\frac{\pi^4}{15}\frac{1}{(4-3q)(3-2q)(2-q)} \quad \text{for } q<1. \tag{A10c}$$

On the other hand the calculation of the integral $I_q(3)$

$$I_q(3)=\int_0^\infty dx \frac{x^3}{[1-(1-q)x]^{\frac{1}{q-1}}-1} \tag{A11}$$

for the case $q>1$ is given below. From the definitions [25]

$$\Gamma(\alpha)\frac{i}{2\pi}\oint_C dv(-v)^{-\frac{q}{q-1}} e^{-v}=1 \tag{A12}$$

where α>0. Substituting $t=v[1-(1-q)x]$ (A13)

one has

$$[1-(1-q)x]^{\frac{1}{q-1}}=\frac{i}{2\pi}\Gamma\left(\frac{q}{q-1}\right)\oint_C dv(-v)^{\frac{-q}{q-1}} e^{-v} e^{-v(q-1)x} \tag{A14}$$

where $\alpha=\frac{q}{q-1}>0$ since $q>1$. On the other hand

$$1=\frac{i}{2\pi}\Gamma\left(\frac{q}{q-1}\right)\oint_C dv(-v)^{\frac{-q}{q-1}} e^{-v}. \tag{A15}$$

If Eq.(A14) is substracted from Eq.(A15) side by side, then one obtains

$$[1-(1-q)x]^{\frac{1}{q-1}}-1=\frac{i}{2\pi}\Gamma\left(\frac{q}{q-1}\right)\oint_C dv(-t)^{\frac{-q}{q-1}} e^{-v}\left(e^{-v(q-1)x}-1\right). \tag{A16}$$

Thus, the substitution of Eq.(A16) into Eq.(A11) leads to

$$I_q(3)=\frac{1}{\Gamma\left(\frac{q}{q-1}\right)\frac{i}{2\pi}\oint_C dv(-v)^{\frac{-q}{q-1}} e^{-v}}\int_0^\infty \frac{x^3}{e^{-v(q-1)x}-1} \tag{A17}$$

or changing the variable of itegration by

$$y=v(1-q)x \tag{A18}$$

Eq.(17) becomes

$$I_q(3) = \frac{1}{\Gamma\left(\frac{q}{q-1}\right)(q-1)^4 \frac{i}{2\pi}\oint_C dv\, (-v)^{-\left(\frac{q}{q-1}-4\right)} e^{-v}} \int_0^\infty \frac{y^3}{e^y - 1} \quad (A19)$$

or

$$I_q(3) = \frac{\Gamma\left(\frac{q}{q-1} - 4\right)}{\Gamma\left(\frac{q}{q-1}\right)(q-1)^4} \cdot \frac{\pi^4}{15} \quad (A20)$$

where from the definition given by Eq.(A12)

$$\frac{1}{\frac{i}{2\pi}\oint_C dv\, (-v)^{-\left(\frac{q}{q-1}-4\right)} e^{-v}} = \Gamma\left(\frac{q}{q-1} - 4\right) \quad (A21)$$

and Eq.(A9a) are used.

In view of the Eq.(A10a); and Eq.(A10b)

$$\Gamma\left(\frac{1}{1-q}\right) = \left(\frac{1}{1-q} - 1\right)\left(\frac{1}{1-q} - 12\right)\left(\frac{1}{1-q} - 3\right)\Gamma\left(\frac{1}{1-q} - 4\right) \quad (A22)$$

is written. Thus, when Eq.(22) is substituted into Eq.(A20) one has

$$I_q(3) = \frac{\pi^4}{15} \frac{1}{(4-3q)(3-2q)(2-q)} \text{ .for q>1} \quad (A23)$$

Finally, one may unify the results which are given by Eqs.(A10c) and Eq.(A23)

$$I_q(3) = \frac{\pi^4}{15} \frac{1}{(4-3q)(3-2q)(2-q)} \begin{array}{l} q<1 \\ q>1 \end{array}. \quad (A24)$$

### Appendix B

In order to calculate the second integral which we face to calculate in Eq.(33)

$$I_q(2) = \int_0^\infty dx \frac{x^2}{[1-(1-q)x]^{\frac{1}{q-1}} - 1} \quad (B1)$$

where q<1 we refer to Eq.(A5). Substitution of the Eq.(A5) into (B1) reads

$$I_q(2) = \frac{\Gamma\left(\dfrac{1}{1-q}\right)}{\int_0^\infty dv\, v^{\frac{1}{q-1}-1} e^{-v}} \int_0^\infty dx\, \frac{x^2}{e^{v(1-q)x}-1} \tag{B2}$$

or changing variable of integration by writing

$$y = v(1-q)x \tag{B3}$$

one obtains

$$I_q(2) = \frac{\Gamma\left(\dfrac{1}{1-q}\right)}{(q-1)^3 \int_0^\infty dv\, v^{\frac{1}{q-1}+2} e^{-v}} \int_0^\infty dy\, \frac{y^2}{e^y - 1} \tag{B4}$$

or

$$I_q(2) = \frac{\Gamma\left(\dfrac{1}{1-q}\right)}{(q-1)^3 \Gamma\left(\dfrac{1}{1-q}+3\right)} \Gamma(3)\zeta(3) \tag{B5}$$

where

$$\int_0^\infty dy\, \frac{y^2}{e^y - 1} = \Gamma(3)\zeta(3) \tag{B6}$$

and

$$\int_0^\infty dv\, v^{\frac{1}{1-q}+2} e^{-v} = \Gamma\left(\dfrac{1}{1-q}+3\right) \tag{B7}$$

are used.

Thus, in view of the Eq.(10a) and Eq.(10 b) which are related to the properties of gamma function, Eq.(B5) reads,

$$I_q(2) = \frac{1}{(3-2q)(2-q)} \Gamma(3)\zeta(3) \tag{B8}$$

where $\Gamma(3)=2!$ ; $\zeta(3)=1.202$.

In order to calculate the integral which is given by Eq.(B1) for the q>1 one may refer to Eq.(A16). Thus, Eq.(B1) reads

$$I_q(2) = \frac{\Gamma\left(\dfrac{1}{1-q}\right)}{\Gamma\left(\dfrac{q}{q-1}\right) \int_0^\infty dv\, (-v)^{-\frac{q}{q-1}} e^{-v}} \int_0^\infty dx\, \frac{x^2}{e^{-v(q-1)x}-1} \tag{B9}$$

where $\alpha = \dfrac{q}{q-1} > 0$ since $q > 1$. If one one changes variable of the integration one obtains;

$$I_q(2) = \frac{\Gamma\left(\dfrac{1}{1-q}-3\right)}{\Gamma\left(\dfrac{q}{q-1}\right)(q-1)^3 \dfrac{i}{2\pi}\displaystyle\int_0^\infty dv(-v)^{-\left(\dfrac{q}{q-1}-3\right)} e^{-v}} \int_0^\infty dy \frac{y^2}{e^y - 1}$$

(B10)

or

$$I_q(2) = \frac{\Gamma\left(\dfrac{1}{1-q}-3\right)}{\Gamma\left(\dfrac{q}{q-1}\right)(q-1)^3}\Gamma(3)\zeta(3) \tag{B11}$$

where from the definition which is given by Eq.(A12) and

$$\frac{1}{\dfrac{i}{2\pi}\displaystyle\int_0^\infty dv(-v)^{-\left(\dfrac{q}{q-1}-3\right)} e^{-v}} = \Gamma\left(\dfrac{q}{q-1}-3\right) \tag{B12}$$

and

$$\int_0^\infty dy \frac{y^2}{e^y - 1} = \Gamma(3)\zeta(3) \tag{B13}$$

are used.

In view of of the properties of the gamma function which are indicated by Eqs.(A10a) and Eq.(10b), one may expresses the term which appears on the right side of Eq.(B12) as

$$\Gamma\left(\frac{1}{1-q}\right) = \left(\frac{1}{1-q}-1\right)\left(\frac{1}{1-q}-12\right)\Gamma\left(\frac{1}{1-q}-3\right) \tag{B14}$$

where $\Gamma(3)=2!$ And $\zeta(3)=1.202$.

Finally by unifying the conclusions of Eqs (B5) and Eq.(B13) one obttains:

$$I_q(2) = \frac{1}{(3q-2)(2q-1)}\Gamma(3)\zeta(3) \quad \begin{array}{l} q>1 \\ q<1 \end{array} \text{ case} \tag{B15}$$

**Acknowledgement**

Fevzi Büyükkılıç and Doğan Demirhan would like to thank Ege University Research Fund for their partial support under the Project Numbers 98 FEN 25.